\documentclass[onecolumn]{article}

\usepackage{color}

\newcommand{\x}{\mathbf{x}}
\newcommand{\C}{\mathbf{C}}
\newcommand{\A}{\mathbf{A}}
\newcommand{\B}{\mathbf{B}}
\newcommand{\W}{\mathbf{W}}
\newcommand{\Q}{\mathbf{Q}}
\newcommand{\bSigma}{\Sigma}
\newcommand{\J}{\mathbf{J}}
\newcommand{\I}{\mathbf{I}}
\newcommand{\uu}{\mathbf{u}}
\newcommand{\G}{\mathbf{G}}

\usepackage[utf8]{inputenc}
\usepackage{geometry}
\usepackage{amsmath}
\usepackage{authblk}
\usepackage{graphicx}
\usepackage{type1cm}
\usepackage{cite}
\usepackage[T1]{fontenc}
\usepackage{caption}

\title{Controllability Analysis of Functional Brain Networks}
\author[a]{Shikuang Deng}
\author[a,b,*]{Shi Gu}%\thanks{Corresponding author: email@mail.com}}}
\affil[a]{School of Computer Science and Engineering, University of Electronic Science and Technology of China, Chengdu, China}
\affil[*]{Corresponding Author.}
\date{}
\begin{document}
\maketitle

\captionsetup[figure]{labelfont={bf},name={Fig.},labelsep=period}

\section*{Abstract}
Network control theory has recently emerged as a promising approach for understanding brain function and dynamics. By operationalizing notions of control theory for brain networks, it offers a fundamental explanation for how brain dynamics may be regulated by structural connectivity. While powerful, the approach does not currently consider other non-structural explanations of brain dynamics. Here we extend the analysis of network controllability by formalizing the evolution of neural signals as a function of effective inter-regional coupling and pairwise signal covariance. We find that functional controllability characterizes a region's impact on the capacity for the whole system to shift between states, and significantly predicts individual difference in performance on cognitively demanding tasks including those task working memory, language, and emotional intelligence. When comparing measurements from functional and structural controllability, we observed consistent relations between average and modal controllability, supporting prior work. In the same comparison, we also observed distinct relations between controllability and synchronizability, reflecting the additional information obtained from functional signals. Our work suggests that network control theory can serve as a systematic analysis tool to understand the energetics of brain state transitions, associated cognitive processes, and subsequent behaviors.

\section*{Introduction}
Large-scale noninvasive neuroimaging provides an accessible window into the rich, complex neurophysiological dynamics of the human brain. Such dynamics are supported by a relatively fixed backbone of white matter fiber bundles spanning cortical and subcortical structures in an intricate network characterized by highly nontrivial topology \cite{hagmann2008mapping,misic2015cooperative}. The relationship between underlying white matter network architecture and large-scale functional dynamics has been the focus of much seminal work, with methods ranging from statistical analyses to biophysical modeling \cite{deco2015rethinking,breakspear2017dynamic}. Yet, across these diverse studies, simple intuitions regarding the mechanisms by which neural activity is propagated along white matter tracts to enable spatially distributed changes in neurophysiological dynamics have been difficult to attain \cite{rosenthal2018mapping,becker2018spectral,medaglia2018functional}. Such difficulties are in part due to the fact that the architecture of the white matter network and the rich dynamics of functional neuroimaging are frequently studied in isolation.

Network control theory is a particularly promising mathematical framework to address these difficulties \cite{liu2011controllability,pasqualetti2014controllability}. Informed by both a precise empirical estimate of white matter network architecture and a model of the dynamics that such an architecture can support, network control theory offers statistics, models, and analytical insights to map and predict the effects of regional activation on time-evolving whole-brain states \cite{gu2015controllability,kim2018role}. Originally developed in the physics and engineering literature \cite{kailath1980linear}, the approach is flexible to applications across scales and species, including cellular models \cite{wiles2017autaptic}, \emph{C. elegans} \cite{yan2017network}, fly \cite{kim2018role}, mouse \cite{kim2018role}, macaque \cite{gu2015controllability}, and human \cite{gu2015controllability}, and has been extended to study cognitive function \cite{cornblath2018sex}, development \cite{tang2017developmental}, heritability \cite{lee2019heritability}, disease \cite{jeganathan2018fronto,bernhardt2019temporal}, and the effects of stimulation \cite{taylor2015optimal,muldoon2016stimulation,medaglia2018network,stiso2019white,khambhati2019functional}. More recently, methods for optimal control have been applied to better understand the mechanisms by which the human brain might switch between diverse cognitive states \cite{betzel2016optimally,gu2017optimal,cornblath2018context,shine2019human}. Despite its broadening utility, current work utilizing notions of network control are somewhat limited by the assumption that all effective relations between regions are time-invariant and encapsulated in the underlying white matter network architecture. Such an assumption leaves the approach agnostic to the distinct ways in which structure can be utilized for inter-regional communication \cite{palmigiano2017flexible}, both to support diverse states in health \cite{cornblath2018context} and in disease \cite{shah2018local}.

Sensitivity to the connectivity elicited by a given state can be partially attained by using methods for the estimation of effective connectivity \cite{friston2011functional}. Common examples of such methods include dynamic causal modeling \cite{friston2011network}, structural equation modeling \cite{mcintosh2012tracing}, Granger causality, and transfer entropy \cite{barnett2009granger}, and can be extended to account for physiological state \cite{havlicek2015physiologically} as well as unknown drivers modulating activity even in quiet resting periods \cite{park2018dynamic}. A limitation of the majority of methods that estimate effective connectivity is that they do not also estimate the dynamics that occur atop that activity. If one wishes to estimate both connectivity and dynamics at once, one naturally turns to the engineering approach of systems identification \cite{zipser1992identification}, a methodology for building mathematical models of dynamic systems using measurements of the system's input and output signals. System identification has been successfully applied to both micro- \cite{mitra2014systems} and macro-circuits \cite{murphy2012isolating}, as well as to human functional magnetic resonance imaging (fMRI) \cite{tauchmanova2008subspace,becker2018large}. While little work has been done in this area, one could naturally consider exploiting system identification \cite{gilson2016estimation} combined with network controllability \cite{zhou1996robust} to investigate the control properties of the brain reflected in functional neuroimaging data.

Here we took exactly this tack using fMRI data from the Human Connectome Project Young Adult 900s release \cite{van2013wu,glasser2013minimal}. Using system identification, we fit the 0- and 1-shift correlation matrices to estimate the system's stochastic linear dynamics \cite{gilson2016estimation}. Using these fitted dynamics, we built a linear control system, setting the transition matrix as the effective connectivity and the control matrix as the canonical form multiplied by the covariance of noise. With this formulation, we asked whether the model dynamics were globally controllable \cite{gu2015controllability,menara2018structured}, and whether the energy required for such control was small or large \cite{klamka1963controllability}. Next, by examining the distribution of minimal control sets and calculating statistics probing two distinct control strategies, we sought to better understand the role of various cognitive systems in regulating whole-brain dynamics. In terms of the additional insight to the structural controllability, we first hypothesize that functional controllability will vary across distinct task states and rest, potentially supporting online adaptation to task demands. We further hypothesize that individual differences in functional controllability can predict task performance. Finally, we examined relations between measurements of functional controllability and of structural controllability to directly assess the value added by the former. Broadly, our study extends current work in network control theory by coupling it with systems identification to better understand the role of effective connectivity in shaping whole-brain dynamics.

\section*{Results}
\begin{figure*}[!htb]
	\begin{center}
		\includegraphics[width=0.85\textwidth]{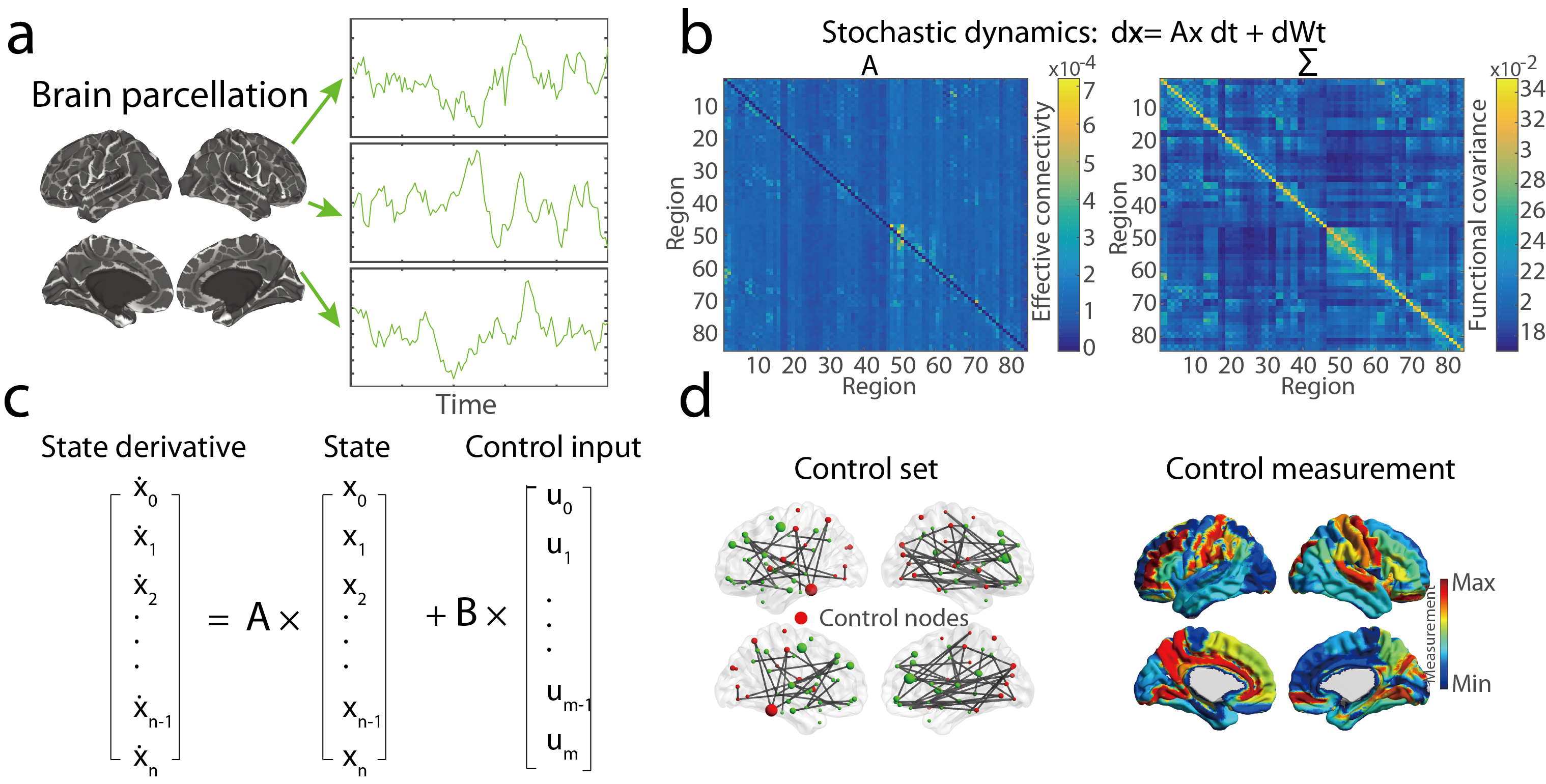}
	\end{center}
	\caption{\textbf{Conceptual Schematic.} \emph{(A)} We begin with the preprocessed BOLD time series from 84 cortical regions. \emph{(B)} We then build a linear stochastic model to represent the BOLD dynamics by estimating the effective connectivity between regions, as well as the covariance of the intrinsic noise. \emph{(C)} We let the intrinsic noise covariance be the control matrix, and we let the effective connectivity be the state transition matrix. This control model preserves the regional activity co-variance pattern estimated from the stochastic model. \emph{(D)} Finally, we detect control sets, quantify controllability statistics, and examine their relations with connectivity.}
	\label{fig:figure1}
\end{figure*}

We used the minimally preprocessed data in the HCP Young Adults 900 released subjects \cite{van2013wu}, which provided {$84\times1200$} (Region $\times$ TR) time series for each of the 758 subjects (Fig.~\ref{fig:figure1} A). We constructed a control dynamics model and estimated the effective connectivity (Fig.~\ref{fig:figure1} B,C). From the effective connectivity, we estimated three control-related statistics: the average controllability, modal controllability, and global synchronizability (Fig.~\ref{fig:figure1} D). Finally, we examined these statistics during the resting state and during the performance of cognitively demanding tasks.

\subsection*{Global Controllability of the System}
First, we sought to address the question of whether the functional brain network is as globally controllable as the structural brain network. Global controllability refers to the capacity to drive a system to any desired state by injecting input into a single node \cite{gu2015controllability}, and can be examined by calculating the smallest eigenvalue of the system's controllability Gramian. We calculated the controllability Gramian of the effective connectivity matrix with perturbations to single-nodes, and found that the smallest eigenvalues ranged from $10^{-42}$ to $10^{-52}$. Consistent with observations in structural networks \cite{menara2018structured}, we concluded that the fitted functional control system is controllable from every region, although the energy required may be large and biological infeasible.

\subsection*{Distribution of Minimal Control Sets}
While global controllability is of theoretical interest, a more practical concern is to identify a set of nodes that can drive desired state transitions with little energy cost. We therefore examined the $\alpha$-minimal control set (Eqn.~\ref{eqn:mcs}) where the weighted adjacency matrix $\A$ was set to be the estimated effective connectivity matrix from Eqn.~\ref{eqn:sde_full}. Here, we set $\alpha = 1$ and identified the 1-minimum control set for each subject; sensitivity and robustness analyses for different values of $\alpha$ can be found in the SI (see Fig.~S1).

\begin{figure*}[!htb]
	\begin{center}
		\includegraphics[width=0.95\textwidth]{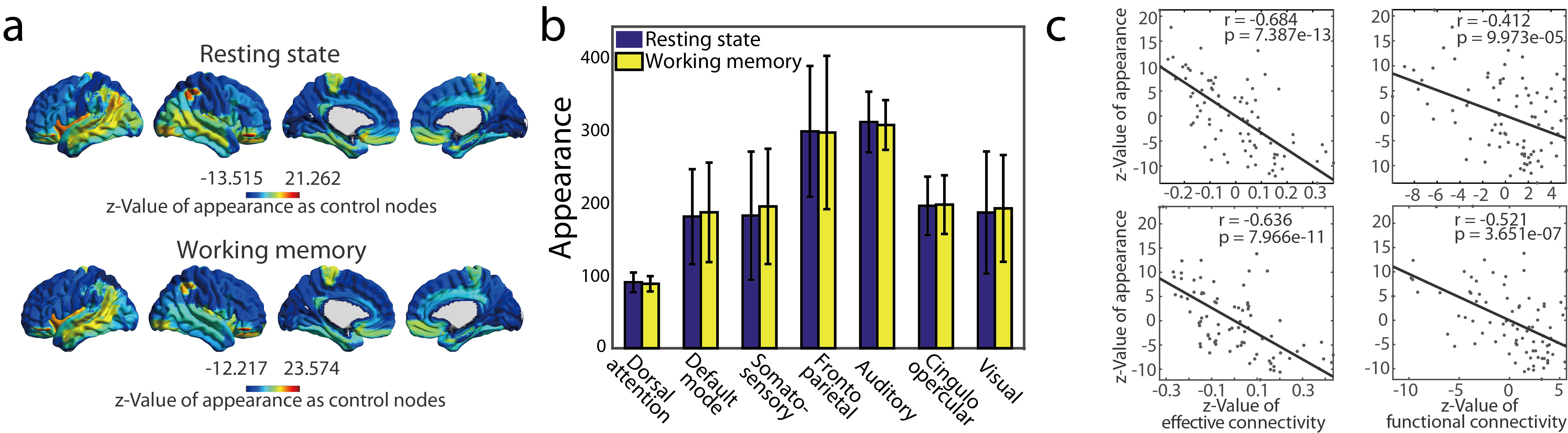}
	\end{center}
	\caption{\textbf{The Spatial Distribution of Minimal Control Sets.} \emph{(A)} Across subjects, the brain regions contained in the 1-minimum control set are spatially distributed, both at rest and during the working memory task condition. \emph{(B)} In the fronto-parietal and auditory systems, we observed a moderately high concentration of control nodes, as measured by the probability of appearance in the 1-minimum control set across subjects. We did not observe significant differences between rest and task conditions in the assignment of control nodes to cognitive systems. \emph{C} We observed that control nodes tend to be located in brain regions that are weakly connected to the rest of the brain, as measured by the z-value of node strength calculated from the functional or effective connectivity matrices. }
	\label{fig:figure2}
\end{figure*}

We first identified the minimal control set for each subject and then calculated the frequency with which each node was found in the control sets of all subjects. By stipulating the null hypothesis that the control set was randomly chosen among all nodes, we calculated the $z$-values of these frequencies with a permutation test. In Fig.~\ref{fig:figure2}A, we showed that the areas most consistently identified as members of the control set were distributed broadly across the brain, including the lateral orbital gyrus, insula, inferior parietal lobule, and middle temporal gyrus.

Next we sought to determine whether the areas consistently identified in the control set tended to be hubs of either functional or effective connectivity networks. We found that the probability of appearing in the minimal control set was negatively correlated with the node strength calculated from either the effective or functional connectivity matrix (Fig.\ref{fig:figure2} C). The finding was consistent across both resting and task conditions. Importantly, this negative relation is intuitive; when controlling a network following the minimal control strategy, control nodes are likely to be weakly connected areas because connections to these nodes could be difficult to cover if control nodes were located far away. Because the control set depends on the lower bound of connectivity, functional brain networks -- which often have many edge weights close to zero -- may be very difficult to control.

\subsection*{Regional Distribution of the Average and Modal Controllability}
After examining the distribution of the nodes in minimal control sets, we further asked how controllable the system was from a given node. For the model setting, we multiplied the square root of the noise covariance matrix to the typical control matrix $\B$ (Equation.~\ref{eqn:lsys}) to take care of the between-region interaction. Next we computed the average controllability, which was defined as the $H_2$-norm of the system, and the modal controllability, which was defined as the inverse-$H_\infty$-norm of the system. We found that the regions with high average controllability were located in precentral \& postcentral, cuneus, temporal suprior, and frontal inferior opercular areas, which were mostly local executive hubs. The regions with high modal controllability were located in olfactory gyrus, and orbitofrontal medial area that were typically thought involved in solving complex tasks. Further, we examined the relationship between controllability and connectivity. We discovered that the average controllability displayed a negative correlation with the effective connectivity strength ($r = -0.721, p = 1.044\times 10^{-14}$) and no significant correlation with functional connectivity strength ($r = 0.035, p = 7.548\times 10^{-1}$) while the modal controllability showed a negative correlation with the functional connectivity strength ($r = -0.751, p = 2.220\times 10^{-16}$) and no significant correlation with the effective connectivity strength ($r = -0.128, p = 2.458\times 10^{-1}$). This supports the claim that the average and modal controllability characterize distinct aspects of the functional dynamics.

\begin{figure}[!htb]
	\begin{center}
		\includegraphics[width=0.85\textwidth]{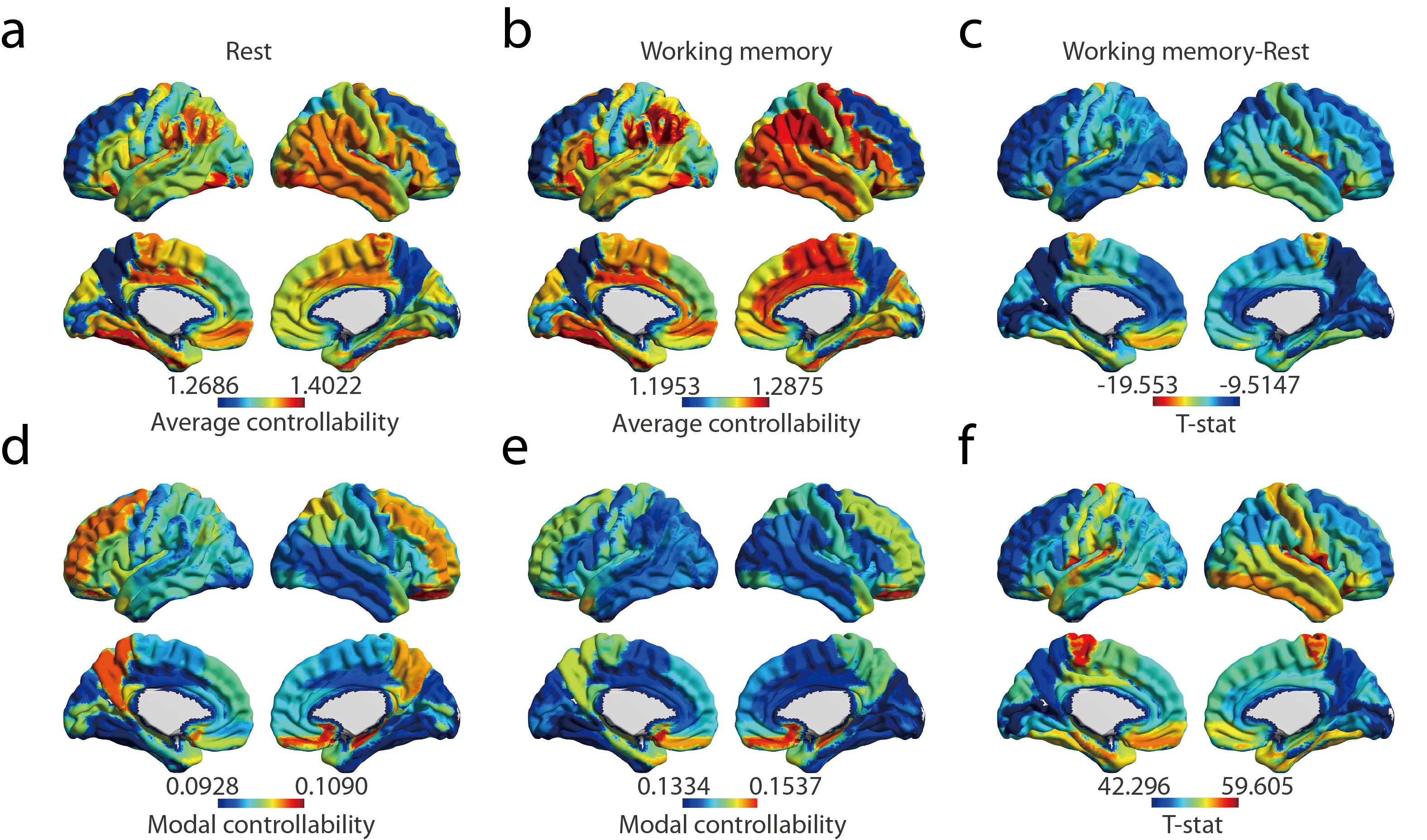}
	\end{center}
	\caption{\textbf{Comparison of Controllability Measurements between the Resting State and the Working Memory Task State.} The average and modal controllability display opposite preference of spatial distribution and maintain a similar pattern for both \emph{(A)(B)} resting state and working memory task state \emph{(C)(D)}. When transitioning from the resting state to the working memory state, \emph{(E)} the average controllability significantly decreases ($FWER < 10^{-5}$) while \emph{(F)} the modal controllability significantly increases ($FWER < 10^{-5}$). The largest differences appear at olfactory cortex and the middle frontal gyrus. }
	\label{fig:figure3}
\end{figure}

\subsection*{Controllability Variation from Resting to Task State}
Controllability was defined through the interaction among regions and related to the energy cost associated with state transitions. From the resting state to the task state, the cognitive control mechanism of the brain altered the dynamics to adapt to the task execution. Here we hypothesized that the average controllability decreased while the model controllability increased from the resting state to task state. Testing with the 2-sample t-test, we found that the mean of average controllability in the memory task was significantly lower than that of the resting state with $t = -13.596, p = 2.232\times10^{-61}$, and the mean of modal controllability on the memory task was significantly higher than that of the resting state, with $t = 52.434, p=1.117\times10^{-115}$. Compared to the structural controllability that focuses on the static connectivity defined through the diffusion, functional controllability allowed for the dependence on observed functional imaging sequences thus better characterized the overall preference of control strategy regarding the underlined states.

\subsection*{Relationship between Controllability and Cognitive Task Performance}
\begin{figure*}[!htb]
	\begin{center}
		\includegraphics[width=0.85\textwidth]{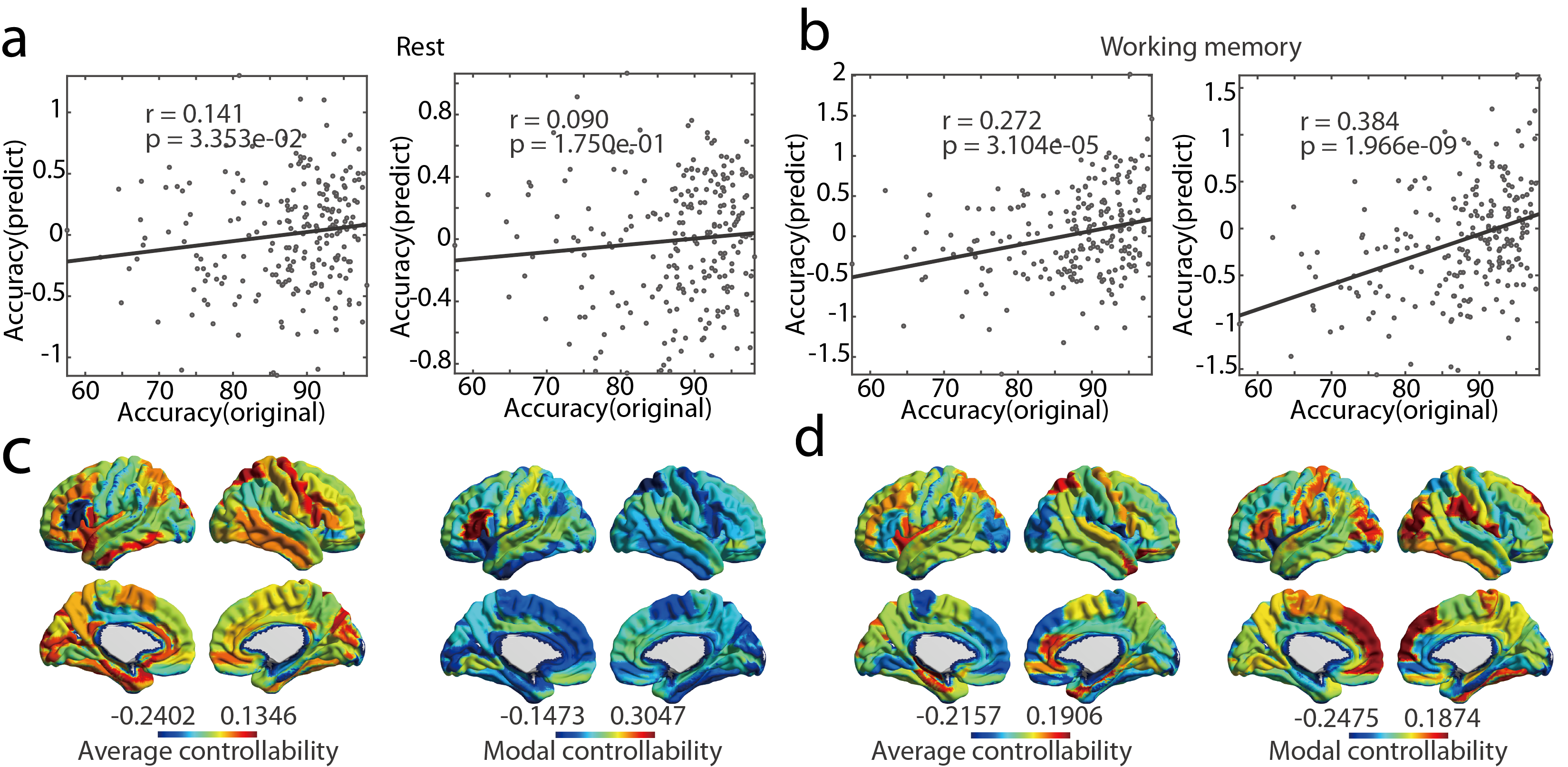}
	\end{center}
	\caption{\textbf{Prediction of Performance in the Working Memory Task from Controllability.} Controllability provides a perspective of understanding the progression of signal and energy across the brain network. Here we want to show that functional controllability can predict the performance in the working memory task. For the resting state, \emph{(A)} the correlation between the true test accuracy and the predicted accuracy from average controllability is $r = 0.141$ with $p = 3.353\times10^{-2}$ and \emph{(B)} the correlation when predicting scores from modal controllability is $r = 0.090$ with $p = 1.750\times 10^{-1}$. For the working memory task state, \emph{(C)} the correlation between the true test accuracy and the predicted accuracy from average controllability is $r = 0.272$ with $p = 3.104\times10^{-5}$ and \emph{(D)} the correlation of modal controllability is $r = 0.384$ with $p = 1.996\times 10^{-9}$. The functional controllability on working memory related areas including the medial frontal gyrus, dorsolateral prefrontal cortex, the supramarginal gyrus and the occipital cortex contribute most positively for the predicted accuracy.}
	\label{fig:figure4}
\end{figure*}

In the previous section, we showed that the functional controllability characterized the transition from resting to task states. Here we further asked whether the individual differences in cognitive performance can be predicted from the functional controllability. To answer this question, we trained a linear model on $70\%$ of the data to predict the score in working memory test with controllability measurement and calculated the correlation between the predicted score and observed score on the left $30\%$ of the data. In Fig.~\ref{fig:figure4}, we first showed that although both the controllability calculated on the resting and working memory task states were able to predict the scores in the working memory task, the correlations were more significant when predicted from the task data ($r = 0.272,p = 3.104\times10^{-5}$ for average controllability and $r = 0.384, p = 1.996\times 10^{-9}$ for modal controllability) than the resting data ($r = 0.141,p = 3.353\times10^{-2}$ for average controllability and $r = 0.090, p = 1.750\times 10^{-1}$ for modal controllability). The predictability of cognitive test performance from controllability measurement holds for the language and emotional intelligence tasks as well. Indeed, the correlation between the predicted and real language task scores were $r = 0.155, p = 0.0196$ from average controllability and $r = 0.212, p = 0.0013$ from modal controllability, and the correlation between the predicted and real emotional intelligence scores were $r = 0.1304, p = 0.0492$ from average controllability and $r = 0.1501, p = 0.0234$ from modal controllability. Next, to examine whether the correlations were significantly different for the resting and testing states, we computed the Fish-z values for comparing two correlations correspondingly and found that both the average controllability ($z=1.45, p = 0.0735$) and the modal controllability ($z= 3.34, p = 4\times10^{-4}$) maintained a more significant prediction with the controllability measurement on the task data than the resting data. In addition, the correlation between observed and predicted accuracy were $r=0.124, p = 6.127\times10^{-2}$ and $r=0.299, p=4.438\times10^{-6}$ when predicted with weighted nodal strength calculated from effective connectivity and the functional connectivity correspondingly (see SI Fig.~S2), lower than those predicted with controllability measurement as well. Specially, the medial frontal gyrus, dorsolateral prefrontal cortex, the supramarginal gyrus and the occipital cortex related to the execution of working memory task appeared as the most positively sensitive areas on the perspective of modal controllability. The temporal pole hippocampus, insula and suproparietal turned out to be the most positively sensitive area on the average controllability. These results support the validity of applying the functional controllability to characterize the state transition from the resting to task states.

\subsection*{Relationship among Controllability Measurements}
\begin{figure*}[!htb]
	\begin{center}
		\includegraphics[width=0.85\textwidth]{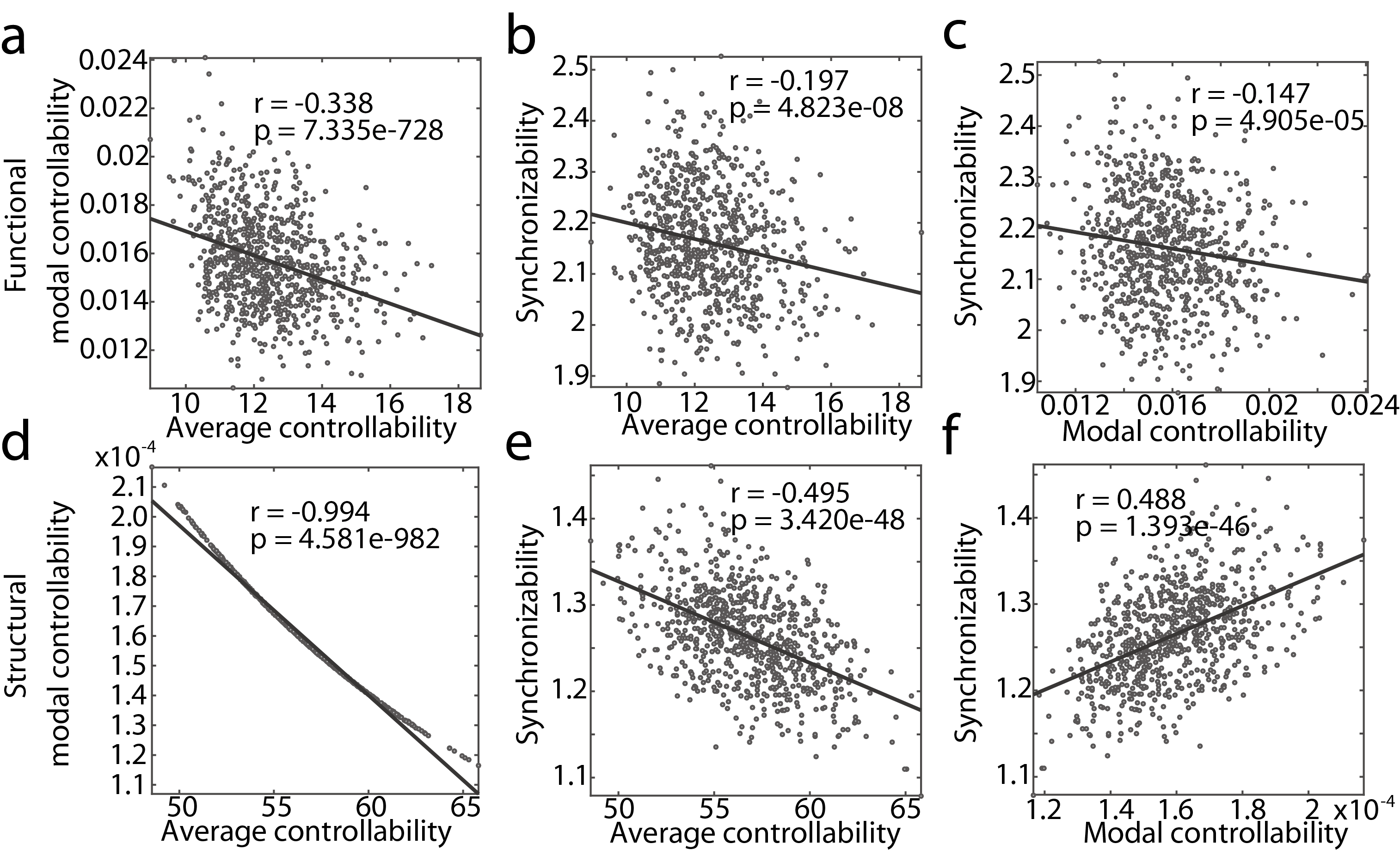}
	\end{center}
	\caption{\textbf{Relationship among Controllability Measurements.} We demonstrate the relationship among the average controllability, modal controllability, and syncronizability when we set the control set as the whole brain. For the functional controllability, the average controllability displays significant  \emph{(A)} negative correlations with modal controllability ($r = -0.338, p = 7.335\times 10^{-728}$ and \emph{(B)} positive correlation with  synchronizability($r = 0.197, p = 4.823\times 10^{-8}$). In addition, \emph{(C)} the modal controllability shows significant negative correlation with synchronizability ($r= -0.147, p = 4.905\times10^{-5}$. For the structural controllability, \emph{(D)}the negative correlation between average controllability and modal controllability is much stronger ($r = -0.994, p = 4.581\times10^{-982}$. \emph{(E)} A similarly stronger negative correlation between average controllability and synchronizability ($r = -0.495, p = 3.420\times10^{-48}$) exists as well. \emph{(F)} However, different from the case of functional controllability, the modal controllability and synchronizability displays a positive correlation ($r=0.488, p = 1.393\times 10^{-46}$.}
	\label{fig:figure5}
\end{figure*}

In the previous section, we investigated the functional controllability and utilized it to investigate the transition from resting to task state. Here we further ask how these measurements were related to each other and how they were different from the established measurements in structural controllability\cite{gu2015controllability}. To answer these questions, we computed the global average and modal controllability by setting the whole brain controlled and correlated them with the global synchronizability. This setting of controlling all nodes could be viewed as an average effect of the controllability across region where the interaction and signal fluctuation happened throughout the whole system. For the calculation of structural controllability adapted from \cite{gu2015controllability}, we normalized the effective connectivity matrix by its largest singular value and used it as the state transition matrix, together with setting identity matrix as the control matrix. From Fig.~\ref{fig:figure5}A, we can see that when all regions were controlled, the average controllability was negatively correlated with the modal controllability for both the functional controllability ($r = -0.338, p = 3.032\times10^{-169}$) and structural controllability ($r = -0.994, p = 4.581\times 10^{-982}$).  The synchronizability, which measured the network's ability of persisting in a synchronous state, displayed a negative correlation with the average controllability in both the functional controllability (Fig.~\ref{fig:figure5}B, $r = -0.197, p = 4.823\times10.^{-8}$) and the structural controllability (Fig.~\ref{fig:figure5}E $r=-0.495, p=3.420\times10^{-48}$). However, the trends were different when correlating synchronizability with modal controllability where the correlation was negative for the functional controllability ($r = -0.0147, p = 4.905\times10^{-5}$) and positive for the structural controllability ($r=0.488, p = 1.393\times10^{-46}$). This could be caused by the fact that in the calculation of structural controllability, only the effective connectivity was adopted with normalization, making the between-controllability measurement somewhat driven by the asymptotic property of the measurement\cite{gu2015controllability} that resulted in higher correlations. While in the functional controllability, both the effective connectivity and noise covariance matrices were involved without extra normalization, which drove the correlation between the two controllability measurements away from the asymptotic behavior thus lower and less significant. One thing to notice was that these relationships were not fully consistent with those for the controllability on structural brain networks through the development\cite{tang2017developmental} (see SI Fig.~S3 for the replication on structural controllability and SI Fig.~S4-6 for the relationship between nodal connectivity strength and controllability), suggesting a mechanical difference between the functional and structural controllability analyses.

\section*{Discussion}
Brain is a complex dynamical system that enables various behaviors through moving itself among multiple cognitive states. Although the trajectories of these state transition are biologically constrained by the white matter microstructure and partially explained by the network control theory based on streamlines\cite{tang2017developmental}, the fitness to the observed dynamics is not satisfying probably due to the intrinsic complexity of modelling the evolving manner of functional signals from the structural connectivity\cite{deco2011emerging,betzel2016optimally,gu2017optimal,gu2018energy}. In practice, it is more feasible yet still critically important to build a control framework from the functional time series, e.g. BOLD, EEG, MEG, and etc with the constraints from structural connectivity constructed from diffusion imaging. In this work, we proposed a novel framework of analyzing brain's functional dynamics from the control perspective where controllability measurements were defined through the system norms and investigated on both the resting and task states.
% point 1: what does functional controllability mean

The controllability on structural brain networks predicts the the ability of alternating large-scale neural circuits based on the assumption that the transition of brain states can be modeled by the structural connectivity\cite{gu2015controllability}. Then, why do we still need the functional controllability? From the opinion of \cite{friston2011functional}, this transition, which quantifies the impact of one region on another in moving the brain states, is indeed one type of effective connectivity. Thus, in addition to obtaining a degenerated model from structural connectivity that dismisses the difference over functional states, it is critically important to model and validate the control theoretical analysis based on effective connectivity. Further, in what space should we model these transitions? The current work of functional controllability examines this problem of defining a set of nodes' role in moving the brain states from a data-driven approach, inferring the transition and co-varying patterns in the signal space. Biologically, the neural stimulus spreads across neurons along the synapse, reflected in the fluctuation of metabolic consumption\cite{kandel2000principles}. Although rigorously the control input should be on the source space\cite{valdes2011effective}, here we focus on the control mechanism and assume that the principle applies in the signal space as well. This assumption is widely supported by the researches that the signal space inference are able to reveal the intrinsic interactions between regions\cite{cohen2017computational,deco2011emerging,fox2005human}. Thus it is meaningful to consider the functional controllability in the signal space.

% explain the shift of the controllability from resting to task
The change of functional controllability from the resting to task state provides a system-control perspective of understanding how the system shifts the association among regions for the adaption to the executive task. We showed that the whole brain's average controllability decreases and the modal controllability increases from the resting state to the task state. This suggests that the resting state is potential `ground state' with better maintain of averaged energy cost. Relatively, for the task state or pluripotent `excited state', more energy would be consumed in order to facilitate the cognitive processes with improved controllability on each mode. This complements the previous reasoning on the regional preference of control strategies\cite{gu2015controllability}. In addition, it unveils how the adjustment of regional activation contributes to the systematic alternation to executing tasks. For the cognitive tasks considered here, we observed larger increase in the task related area and smaller decrease in the default mode area. This indicates that the compensation on the task state is lower compared to other regions, backing the previous literature stating that the default mode is probably optimized for the baseline state. Further, this result enriches the reasoning on the relationship between weakly connected area and modal controllability. The modal controllability quantifies a region's ability in controlling the amplitude of signal amplifying thus it is higher on the weakly connected area because the amplifying effect was transitioned through the connectivity, making the less connected region result in higher controllability of amplifying amplitude. When compared through the resting and task states, the increase of modal controllability on the corresponding regions improves the controllability on its associated modes thus the execution of task as well\cite{cui2018optimization}.

% relation to cognition and psychiatry works done by fMRI
Controllability not only characterizes the shift of states from resting to task states but also predicts the individual difference in the scores in the working memory tasks. Why? On the one hand, as shown in previous literature\cite{finn2015functional,shen2017using}, the functional connectome encodes the information of individual difference. On the other hand, from the perspective of system control, higher modal controllability indicates more efficient utility in executing the mode thus the coefficients on the task-related area would contribute positively in the predictive model\cite{medaglia2018network,cornblath2019sex}. The connectivity-based models, although informative about the spatial location of the regions related to the task performance, lack a mechanical explanation of predictability, which is complemented by the proposed controllability framework.

% point 2: compare functional and structural controllability
The controllability of functional brain networks is closely related to that of structural networks. Both frameworks relied on the time-invariant linear model, which could be not real considering the nonlinearity of brain system yet still provides a fair estimation both from the perspective of behavior and control theory. Second, the average controllability display strong positive correlations with nodal connectivity strength in the structural networks and with effective connectivity strength in the functional networks. This is supported by the results revealed in \cite{deco2011emerging} that the structural connectivity can act as a predictor of the effective connectivity, which also provides the consistency between the current framework and the controllability analysis on structural brain networks\cite{gu2015controllability}. In addition, this positive correlation suggested that although not exactly overlapped, the structural and functional hubs maintain efficient roles in driving the brain to many easily-reachable states, providing an explanation of the cognitive association for both structural and functional hubs.

However, the differences also exist between the two proposed networks. On the modeling perspectives, as the functional control model fits the BOLD time series directly while the structural control model studies the induced dynamics, the two frameworks are generally only applicable to their own modalities. The newly proposed one does have some convenience. For example, the controllability frameworks on the structural networks requires the normalization of transition matrix to ensure the Schur stability of the system while the proposed framework on functional networks satisfied the constraints automatically from the model fitting. This avoids the potential bias introduced by the normalization when the controllability statistics need to be averaged across subjects\cite{wu2018benchmarking}. In addition, the definition of controllability on structural brain networks was derived from the connectivity\cite{pasqualetti2014controllability,gu2015controllability} and regionally defined while the currently proposed framework is defined via the system norms that explicitly links the connectivity to a formulated energy and naturally extendable to control sets consisting of multiple and even all nodes. On the application scenario, structural controllability framework provides a mechanical explanation of how the underlined structure supports the executive function\cite{cornblath2019sex} and neural development\cite{tang2017developmental}, as well as the evolution of dynamic trajectories associated with the state transition\cite{gu2017optimal}. Yet it remains to explain how the biomarkers defined through the activation of and the statistical association among regions are related to the system's controllability from a mechanical view. For example, different brain activation and connectivity patterns act as biomarkers to unveil the representative phenotype of psychiatric disease like depression\cite{anand2005activity,veer2010whole} and schizophrenia\cite{lynall2010functional,fitzsimmons2013review}. Yet it is unclear how one or multiple regions, i.e. the control set, drive the whole neural circuit to move across states and result in the abnormality in brain functions. Our functional controllability model potentially bridges the gap by analyzing the time-series directly rather than inferring from the structure\cite{bernhardt2019temporal}. It allows future possibility of application to intervene the neural circuits via certain nodes for psychiatric medication\cite{sitaram2017closed}.

Methodologically, it is worth pointing that the current effort still shares certain limitations before. First, by changing the dynamics into a stochastic one, the nonlinear effect still remained to be solved in future. Secondly, due to the constrains of unpredictable noise in the measurement, the approximation to the observed trajectories are still unsatisfactory especially for the real values. Finally, the controllability measurements are highly related to the effective and functional node strength. Although the linear dynamics could predict the trend as we show in the article, quantitatively, the amount of modeled dynamics is still around the limit point of the linear system thus may not be able to quantify the long-range high level dependence on connectivity.

\section*{Materials and Methods}
The theory in this work consists of two parts. The first part is the inference on dynamics, which assumes that the dynamical patterns of BOLD signals are driven by the intrinsic fluctuation of noise and can be reflected on its covariance structure. The second part is how a region imposes its impact on others, i.e. the definition of controllability. We denote the state at time $t$ for a brain as $\x_t$, which is an $N\times1$ vector with $N$ as the number of regions. Usually, the evolutionary dynamics of the states is formulated as describing the state's time derivative $dx/dt$ with the state variable $x$ and other parameterized related terms, e.g. the noise. In this work, we attempt to  fit the dynamics of $\x_t$, followed by investigating it from the control perspectives where we examine both the spatial distribution of minimal control sets and control measurements.

\subsection*{Preprocessing of fMRI Data}
We used the minimally preprocessed fMRI data conducted using HCP Functional Pipeline v2.0 \cite{glasser2013minimal}. Subjects with incomplete resting state or two task data were excluded. Then we used DPARSF \cite{yan2016dpabi} and SPM12 \cite{ashburner2014spm12} to process these minimally preprocessed data. First, we removed the constant, linear and quadratic trend from these functional images. Next, several nuisance signals  including cerebrospinal fluid signal, white matter signal, and motion effect were regressed from the time course of each voxel using multiple linear regression and Friston's 24 head motion parameters. Then 3D spatial smoothing was applied to each volume of the fMRI data using a Gaussian kernel with Full-width at Half Maximum (FWHM) equaling to 4 mm. Finally, ALFF and fALFF(0.01-0.1 Hz) was used to inhibit the energy of physiologically meaningless brain regions, and temporal band-pass filtering (0.01-0.1 Hz) was applied to reduce the influence of low-frequency drift and the high-frequency physiological noise. For resting state fMRI time series, the first and last 50 volumes were discarded to suppress equilibration effects. For task data, the break time in the task was deleted to remove the effects of resting state.The AAL2 atlas \cite{rolls2015implementation} was used for the parcellation of brain cortex into region. We kept 84 cortical regions only excluding non-cortical regions including amygdala, caudate, putamen, pallidum, thalamus, vermis and cerebellum areas. Finally there are 758 subjects used in the current analyses, including 422 females and 336 males aged from 22 to 37 years old.

\subsection*{Construction of the Control Dynamics}
We start from the linear stochastic model, where the changing rate of the state is determined by the current state and the random diffusion following Gilson' s steps\cite{gilson2016estimation}. Mathematically, the dynamic model is given by
\begin{equation}
\label{eqn:sde_full}
	d\x= (-\frac{1}{\tau_x}\x + \C\x )dt + d\W_t,
\end{equation}
where $\tau_x$ is the constant of state decay over time, $\C$ is the effective connectivity matrix and $d\W$ is a wiener process with covariance $\bSigma$. To fit the dynamics, we estimate three unknown parameters, $\tau_x$, $\C$ and $\bSigma$ by minimizing the loss between model-derived and empirical covariances. First, assuming the stationarity of this system, we can derive the relationship between autocorrelation $Q^0$ and covariance $\bSigma$ with the Ito's formula\cite{ito1973stochastic} which implies
\begin{equation}
	\J\Q^0 +\Q^0 \J^\dagger +\bSigma = 0,
\end{equation}
where $\J = -\frac{1}{\tau_x}\I +\C$ is the Jacobian of equation \ref{eqn:sde_full}. Further, the theoretical formula of the $\tau$-delay autocorrelation can be computed as $\Q^{\tau} = \Q^0 \exp(\J^\dagger \tau)$ via similar derivation. On the other hand, we can define the empirical estimation of autocorrelation $\hat\Q^k$ with k-shift. We hope the fitted autocorrelations are as close to the empirical ones as possible. Thus the loss function of fitting the dynamics is given by the weighted sum of the distance between each pair of estimated and empirical auto-covariance matrices, i.e.
\begin{equation}
	\mathcal{L}(\J,\bSigma) = \sum_{k=1}^K\lambda^k l(\Q^k(\J,\bSigma) , \hat\Q^k),
\end{equation}
where $l(\cdot)$ is the loss function between two covariance matrices, $K$ is the number of shifts we want to use for the estimation and $\lambda$ is a scalar to weight among the losses for these $\Q^k$'s.

Using the gradient descent, $\hat{J},\hat{\bSigma}$ can be recursively updated.

Analogous to the classical control representation $\dot{x}(t) = \A \x(t) + \B \uu(t)$, the state transition matrix $\A$ can be modeled with the Jacobian $\J$ and the control input matrix $\B$ is reformulated as $\hat{\bSigma}\cdot\B_\mathcal{K}$, which simultaneously selects the control sets with $N\times K$ matrix $B_\mathcal{K}$ and preserves the co-varying pattern estimated from the stochastic modeling with $\hat{\bSigma}$. Consequently, we built up the linear time-invariant dynamic model for brain's functional signals as
\begin{equation}
	\frac{d\x}{dt} = \hat{\J}\x + \hat{\bSigma}\cdot\B_\mathcal{K}\uu(t),
	\label{eqn:lsys}
\end{equation}
where $\uu(t)$ is the input vector to be determined. In this manuscript, for the ease of notation, we use $\A=\hat{\J}$ and $\B=\hat{\bSigma}\cdot\B_\mathcal{K}$ when there is no ambiguity. When we say functional connectivity, we refer to the Pearson's correlation for the time series of each pair of regions.

\subsection*{Identification of Minimal Control Sets}
Theoretically, if the transition matrix $\A$ for the linear system is non-degenerate, the system is almost surely controllable from a single node\cite{menara2017number}. But the control energy could be so high that results in unreasonable trajectories in practice. Here we adapt the minimum dominant set algorithm in \cite{sun2017understanding} and define the $\alpha$-minimum control set ($\alpha$-MCS) as the solution of the following optimization problem:
\begin{equation}
	\label{eqn:mcs}
	\begin{aligned}
		& \underset{\mathcal{K}}{\text{min}}
		& &  \sum_{k\in \mathcal{K}}\beta_k,\\
		& \text{s.t.}
		& & \sum_{k\in \mathcal{K},k\neq i} \beta_k a_{ik}  \geq \alpha\cdot  (1-\beta_i)\cdot\text{max}(\A),\;
	\end{aligned}
\end{equation}
where $\A=\{a_{ij}\}$ is the transition matrix, $\beta_i$ takes 1 if node $i$ is chosen as a control node and 0 otherwise, and $\mathcal{K}$ is the control set. When the network is binary and $\alpha = 1$, it reduces to the regular problem of identifying the minimal control set. When $\A$ is weighted, the optimization problem finds the minimal set such that every nodes is either in the control set or connected to the control set with overall strength above a threshold $\alpha$ scaled by the maximum weight in the weighted adjacency matrix.

\subsection*{Average Controllability}
The average controllability of the linear stable system refers to its $H_2$-norm, which intuitively quantifies the average distance the system can reach in the state space with unit input energy. Mathematically $H_2$ norm is the energy of the output of the system
\begin{equation}
	\dot{\x}  = \A \x +\sum_i B_i \boldsymbol{\omega_i}
\end{equation}
where $\omega_i = \delta_i(t)$ is the $\delta$-function and $B_i$ is the i-th column control matrix in Eqn[\ref{eqn:lsys}]. The average controllability is then defined as
\begin{equation}
	a_c = H_2 = \sqrt{\mathrm{trace}\left[\B^T\left(\int_{0}^{+\infty}\exp{(\A t+\A^Tt)}dt \right)\B\right]}
	\label{eqn:ac}
\end{equation}
where $\B$ is the control matrix. If the average controllability is high, it means that the brain is more efficient in moving into many easily reachable states.
\subsection*{Modal Controllability}
The modal controllability of the linear stable system is defined as the inverse of $H_\infty$-norm, which quantifies the inverse of maximal possible vector amplification with $\sin(\cdot)$ input. Mathematically, it is defined as
\begin{equation}
	g_c = (H_\infty)^{-1} = \left(\sup_{\omega\in\mathbf{R}}{\sigma}\{\G(j\omega)\}\right)^{-1}
	\label{eqn:gc}
\end{equation}
where $j$ is the virtual unit with $j^2 = -1$, $\G(s) = (s\I-\A)^{-1}\B$, and ${\sigma}$ denotes the largest singular value. A higher modal controllability then corresponds to a easier control of the dynamics in the direction of highest energy cost.
\subsection*{Global Synchronizability}
The global syncronizability refers to the inversed spread of the Laplacian eigenvalues, which intuitively measures the ability of the network's dynamics to persist in a synchronous state where all nodes have the same magnitude of activity\cite{tang2017developmental}. Mathematically, it is defined as
\begin{equation}
	s_c = \sqrt{\frac{d^2(N-1)}{\sum_{i=1}^{N-1}|\lambda_i - \overline{\lambda}|^2}}
	\label{eqn:sc}
\end{equation}
where $\lambda_i$ is the positive eigenvalues of the Laplacian matrix $\mathbf{L}$ with $L_{ij} = \delta_{ij}\sum_k A_{ik} - A_{ij}$ and $d = \sum_i\sum_{j\neq i} A_{ij} /N$ is the average strength of each node.
\subsection*{Prediction with Linear Model}
To examine the effectiveness of nodal measurement to predict the performance in memory task, we build a linear model with the nodal measurement $\mathbf{z}=\{z_i\}$(e.g. controllability measurement or weighted nodal strength) as input and the task score $t$ as the output. Mathematically, we write the model as $t \sim \boldsymbol{\eta}\cdot\mathbf{z}$, where $\boldsymbol{\eta}=\{\eta_i\}$ represents the contribution of each region's nodal measurement in predicting the score in cognitive tasks.
\section*{Acknowledgments}
This work is primarily supported by NSFC 61876032.
\section*{Author contributions}
S.G. designed the research. S.G. and D.S.K performed the research, contributed to new analytical tools and wrote the draft.
\section*{Competing Financial Interests}
The authors declare no competing financial interests.

% the limitation of current work
% point 3: relationship with structural controllability
% point 4: relationship with cognitive control
% point 5: what's the potential advantage and how it can contribute for the multi-modality analysis
% point 6: limitation

%\bibliographystyle{plain}
%\bibliography{test}
\end{document}